\documentstyle[epsfig,twoside,12pt]{article}
\def\greaterthansquiggle{\raise.3ex\hbox{$>$\kern-.75em\lower1ex\hbox{$\sim$}}}
\def\lessthansquiggle{\raise.3ex\hbox{$<$\kern-.75em\lower1ex\hbox{$\sim$}}}

\newcommand{\beq}{\begin{equation}}
\newcommand{\eeq}{\end{equation}}
\newcommand{\beqa}{\begin{eqnarray}}
\newcommand{\eeqa}{\end{eqnarray}}
\newcommand{\beqan}{\begin{eqnarray*}}
\newcommand{\eeqan}{\end{eqnarray*}}
\newcommand{\ba}{\begin{array}}
\newcommand{\ea}{\end{array}}

\newcommand{\sh}{{\rm sh}}
\newcommand{\Det}{{\rm Det}\,}

\newcommand{\ra}{\rightarrow}

\newcommand{\ve}{\varepsilon}

\newcommand{\A}{{\cal A}}

\newcommand{\cO}{{\cal O}}

\newcommand{\R}{{\cal R}}
\newcommand{\cS}{{\cal S}}

\newcommand{\dfrac}{\displaystyle \frac}
\newcommand{\dint}{\displaystyle \int}
\newcommand{\dsum}{\displaystyle \sum}
\newcommand{\dprod}{\displaystyle \prod}

\def\nz{\ifmmode {I\hskip -3pt N} \else {\hbox {$I\hskip -3pt N$}}\fi}
\def\zz{\ifmmode {Z\hskip -4.8pt Z} \else
       {\hbox {$Z\hskip -4.8pt Z$}}\fi}
\def\qz{\ifmmode {Q\hskip -5.0pt\vrule height6.0pt depth 0pt
       \hskip 6pt} \else {\hbox
       {$Q\hskip -5.0pt\vrule height6.0pt depth 0pt\hskip 6pt$}}\fi}
\def\rz{\ifmmode {I\hskip -3pt R} \else {\hbox {$I\hskip -3pt R$}}\fi}
\def\cz{\ifmmode {C\hskip -4.8pt\vrule height5.8pt\hskip 6.3pt} \else
       {\hbox {$C\hskip -4.8pt\vrule height5.8pt\hskip 6.3pt$}}\fi}

\def\au{{\setbox0=\hbox{\lower1.36775ex%
\hbox{''}\kern-.05em}\dp0=.36775ex\hskip0pt\box0}}
\def\ao{{}\kern-.10em\hbox{``}}

\normalsize
\begin{document}
\bibliographystyle{plain}
\begin{titlepage}
\begin{flushright}
UWThPh-2000-37\\
ESI-953 (2000)\\
math-ph/0010038\\
October 25, 2000
\end{flushright}

\vspace*{2.2cm}
\begin{center}
{\Large\bf  Second-quantization picture\\[5pt]
of the edge currents in the\\[8pt]
fractional quantum Hall effect$^\star$}\\[36pt]

Nevena Ilieva$^{\ast,\sharp}$  and Walter Thirring  \\ [12pt]
{\small\it
Institut f\"ur Theoretische Physik \\ Universit\"at Wien\\
%Boltzmanngasse 5, A-1090 Wien \\
\smallskip
and \\
\smallskip
Erwin Schr\"odinger International Institute\\
for Mathematical Physics}\\

\vspace{0.8cm}
\begin{abstract}
We study the quantum theory of two-dimensional electrons in a
magnetic field and an electric field generated by a homogeneous
background. The dynamics separates into a microscopic and a
macroscopic mode. The latter is a circular Hall current which is
described by a chiral quantum field theory. It is shown how in
this second quantized picture a Laughlin-type wave function
emerges.

\vspace{0.8cm}

\hspace{1.9cm}Laughlin wave function

\end{abstract}
\end{center}

\vfill
{\footnotesize

$^\star$ Work supported in part by ``Fonds zur F\"orderung der
wissenschaftlichen Forschung in \"Osterreich" under grant P11287--PHY;

$^\ast$ On leave from Institute for Nuclear Research and Nuclear
Energy,
Bulgarian Academy of Sciences, Boul.Tzarigradsko Chaussee 72, 1784 Sofia,
Bulgaria

$^\sharp$ E--mail address: ilieva@ap.univie.ac.at
}

\end{titlepage}

\setcounter{page}{2}
\section{Introduction}

The discovery of the fractional quantum Hall effect (FQHE) \cite{Nob, L1}
marked a new era in condensed matter physics, both theoretical and
experimental. This effect takes place in two-dimensional electron
systems in a strong magnetic field. It occurs because the (Coulomb)
electron-electron interaction results in the formation of highly
correlated incompressible states \cite{buch}, despite the fact that the
lowest Landau level is only partially filled. The electron systems which
demonstrate a FQHE (and are called FQH liquids) in fact represent a
whole new state of matter. For its description one has to completely
abandon the theories based on the single-body picture (such as the
Fermi-liquid theory) but use an intrinsic many-body theory, e.g. the one
proposed by Laughlin \cite{L1} and develope adequate new
techniques and concepts (as the one of {\it topological order\/}
\cite{Wen}).

All bulk excitations of the FQH liquids have finite positive energy gap.
With the gauge arguments in Refs. \cite{L2, H, W1} one gets
convinced that the FQH states should also support gapless edge
excitation, similarly to the IQH case, but which, contrary to the latter,
cannot be described by a chiral 1D-Fermi liquid theory.  In the case at
hand,
the topological nature of the large-scale physics of the FQH liquids
provides an effective-theory description of their bulk properties by
means of a topological Chern--Simons theory \cite{FK, FZ, W1}. In
particular, based on the connection between the three-dimensional
topological Chern--Simons theory and the two-dimensional chiral
Wess--Zumino--Witten model (Kac--Moody algebra), discovered
by Witten \cite{Wit} and constructively developed in \cite{FKi},  it has
been realized \cite{W2, FK} that  the edge currents of an arbitrary QH
fluid in an incompressible state generate a chiral current (Kac--Moody)
algebra. This observation suggested the application of methods from
chiral conformal field theory to the analysis of incompressible QH liquids.
In this relation one traces the now-a-days popular {\it holographic
principle\/} \cite{BS}. In our case it states that the topological field
theory describing the scaling limit of the bulk of an incompressible QH
fluid is completely determined by a chiral conformal field theory
describing the edge degrees of freedom of such a fluid
with the same Hall conductivity \cite{FP}.

In general, the theory of the edge degrees of freedom is more complicated
and less universal than the theory of the bulk \cite{ITT}. Still, the edge
excitations which form the so-called chiral Luttinger liquid (CLL) provide
us a practical way to measure topological orders in experiment. However,
most of the considerations of the edge states rest upon an effective-theory
analysis (though in \cite{Sch} a reformulation of the edge theory directly
in terms of a set of fundamental excitations has been attempted) and we
will strive for consistent derivation from a microscopic theory.

The quantum one-dimensional anyon fields (in particular, the noncanonical
fermions) constructed in \cite{epj, tmp, INT}, are a reasonable candidate
for this role.  It is our purpose in the present note to show how do they
originate from the initial two-dimensional Fermi algebra, what type of states
do they form  and how does the whole picture change with the temperature.
In addition, one point will be clarified. Recall, that as specially emphasized
by Haldane \cite{Hal}, the key step in Laughlin's treatment of the FQHE
has been to abandon conventional second-quantized methods, which had
proved fruitless, and return to a first-quantized description. The
noncanonical fermions in question in fact relate the first and second
quantized pictures of the FQHE.

More in detail, we shall consider electrons in a plane with a constant
magnetic field perpendicular to it and an electric field generated by a
homogeneous background harmonic potential. When only a magnetic
field present, the one-particle observables form two independent (mutually
commuting) canonical pairs
--- the velocities and the centers of Larmor orbits. Correspondingly, the
Hilbert space of the (first) quantized theory has a tensor-product structure
(see also \cite{KL}).  In general, an arbitrary electric field would spoil it, but
it turns out that this does not happen for the particular (radial) electric field
we have chosen. In this case the time evolution respects this product
structure and factorizes into a microscopic and a macroscopic motion. Also
upon second quantization we have a tensor product,  the second factor
corresponding to the edge currents mentioned above. In the thermodynamic
limit one obtains a $(1\!+\!1)$-dimensional chiral quantum field with the
possibility to use all results that are available for it.

\section{Preliminaries}

We consider the motion of electrons in two dimensions in a constant
magnetic field $\bf {B}$ perpendicular to the plane of motion and an
electric field $\bf {E}(\bf {x})$ generated by a homogeneous background
charge. In units $e=m=1$ the one-particle Hamiltonian is
\beq
H = {1\over2}\left[\left(p_1 + {B x_2\over2}\right)^2 +
\left(p_2 - {B x_1\over2}\right)^2\right] + {E\over2}(x_1^2+x_2^2)\,,
\quad B, E>0
\eeq
such that ${\bf E}({\bf x})=E\,{\bf x}$ and $B = |{\bf B}|$. Since for many
particles the Coulomb repulsion cannot be treated exactly one might think
that it is to some extent taken care of by a partial neutralization
of the background and we consider the case $B^2 \gg E$.

The classical motion has a high-frequency mode corresponding to the
cyclotron circles in $B$ and a low-frequency rotation in the opposite
direction of the centers of these circles showing the Hall effect
generated by $\bf  {E}$, Fig.1.
%%%%%%%%%%%%%%%%%%%%%%%%%%%%%%%%%%%%%%%%%% 
%%%%%%Fig.
%\includegraphics*[110,20][350,220]{tri.eps}
\begin{figure}[ht]
\setlength{\unitlength}{1pt}
\begin{picture}(240,210)
\put(130,-20){\makebox(0,0)[lb]
{\scalebox{0.8}%
{\includegraphics*[110,5][350,210]{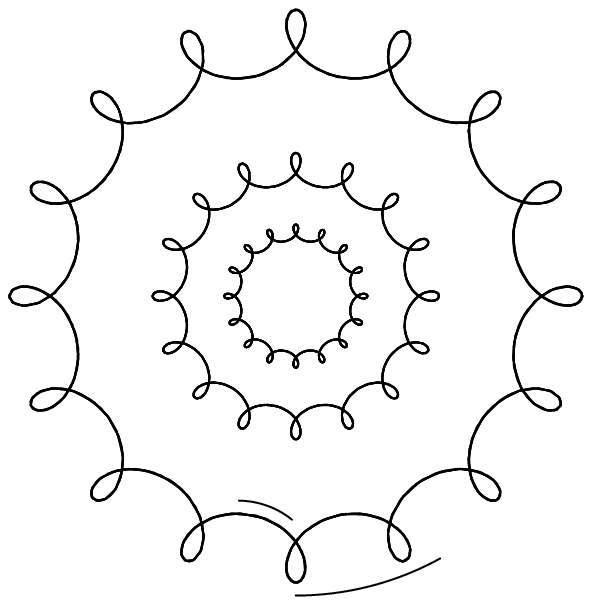}}}}
\put(212,28.5){\makebox(0,0)[l]{{\tiny $<$}}}
\put(223,7){\makebox(0,0)[l]{{\tiny $<$}}}%\prec
\put(275,20){\makebox(0,0)[l]{{\small $\omega_H={B-b\over 2}$}}}
\put(118,35){\makebox(0,0)[l]{{\small $\omega_c={B+b\over2}$}}}
\end{picture}
\caption{{\it Classical Larmor precession in harmonic background.}}
\end{figure}
%%%%%%%%%%%%%%%%%%%%%%%%%%%%%%%%%%%%%%%%%%
To separate these modes we  recall that in a
magnetic field velocity components provide a pair of canonical variables
\beq
{\bf v} = \left(p_1 + B x_2/2 , p_2 - B x_1/2 \right) =  (q, p)\,, \quad
\{q, p\}=B\,.
\eeq
Another (independent) canonical pair is given by coordinates of the
centers of the cyclotron circles
\beq
\bar {\bf x} = (x_1/2 + p_2/B, x_2/2 - p_1/B) = (\bar q, -\bar p)\,,\quad
\{\bar q, \bar p\} = 1/B\,.
\eeq
In these variables Hamiltonian (1) separates into two oscillators
with an effective magnetic field $b = \sqrt {B^2+4E}$ that is induced
by the electric field:
\beq
H = [(1 + B/b){\bf v}_b^2 + b(b-B)\bar {\bf x}_b^2]/4\,,
\eeq
where in ${\bf v}_b$, $\bar {\bf x}_b$ (as well as in $q, p, \bar q, \bar p$
below)  $B$ is replaced by $b$.

For the complex coordinates
\beq
\ba{lcl}
a = (q + ip)/\sqrt{2b}\,, & \quad & c = (\bar q + i\bar p)\sqrt{b/2}\\[6pt]
a^* = (q - ip)/\sqrt{2b}  \,, & \quad & c^* = (\bar q - i\bar p)\sqrt{b/2}
\ea
\eeq
the Poisson brackets are
\beq
\{a^*, a\} = i\,, \quad \{a,\hat L\} = ia
\eeq
$$
\{c^*,c\} = i, \quad \{c^*,\hat L\} = ic^*, \quad \{c,a\} = \{c,a^*\} = 0
$$
with
\beq
\hat L = x_1p_2 - x_2p_1
\eeq
being the generator of rotations. This makes $a$ the high-frequency mode,
\beq
a(t) = e^{-it(b+B)/2}a(0)\,, \quad \hat L(t) = \hat L(0)\,.
\eeq
whereas $c$ shows a low-frequency rotation
\beq
c(t) = e^{-iv t}c(0)\,, \quad v = (b-B)/2\,.
\eeq
In the limit considered $v\ra E/B$ and we get the usual Hall velocity
${\bf v} = {\bf B}\times {\bf E}(x)/|{\bf B}|^2$.

In quantum theory the eigenvalues of ${\bf v}_b^2/2$ and
$\bar{\bf x}_b^2/2$ are $b(n+{1\over 2})$, resp. $(m+{1\over 2})/b$ so
that the spectrum of $H$ becomes
\beq
E_{n,m} - E_0 = n(b+B)/2 + m(b-B)/2
\eeq
Upon first quantization $a, \hat L, c$ become operators which satisfy (6)
with $\{\,,\,\} \ra -i[\,,\,]$. The time evolution $a(t) = e^{iHt}a(0)e^{-iHt}$
and similarly for $\hat L, c$ remains the same, Eqs.(8), (9).

For the ground state $\Psi_0$, $(H-E_0)\Psi_0 = 0$ we must have
\beq
a\Psi_0 = \hat L\Psi_0 = 0\,.
\eeq
Furthermore, $a^*$ and $c$ decrease $\hat L$ by $1$ so we must also
have $c\Psi_0 = 0$ since by (1) $H\geq 0$. The eigenstates of $H$,
which correspond to the eigenvalues $E_{n,m}$ from (10) are thus
constructed in the usual manner
\beq
\Psi_{n,m} = \frac{c^{*m} a^{*n}\Psi_0}{\sqrt{n!\,m!}}
\eeq
and are simultaneously eigenstates of the angular momentum $\hat L$.

\paragraph{Remarks}
\begin{enumerate}
\item $n$ labels the Landau levels and $m$ shows how their degeneracy is
lifted by the electric field $E$ of the background (observe $0<v<E/B$).
\item $\Psi_{n,m}$ as a function of $(x_1,x_2)$ can easily be given as in
\cite{L1} and in our notations reads
$$
\ba{ccl}
\Psi_{n,m} & = & (\pi\,m!\,n!\,2^{m+n+1}\,b^{m+n-1})^{-1/2}\,
e^{\,b(x_1^2+x_2^2)/4}\\[6pt]
& \times & \left({\partial\over\partial x_1} - i {\partial\over\partial x_2}\right)^n\,
\left({\partial\over\partial x_1} + i {\partial\over\partial x_2}\right)^m\,
e^{-b(x_1^2+x_2^2)/2}\,.
\ea
$$
\item The particle density in the lowest Landau level is proportional to the
effective magnetic field $b$
$$
\Psi_{0,m} = \sqrt{b/2\pi\,m!}\, (b/2)^{m/2} z^m\, e^{-b\bar zz/2}\,,
$$
$$
\sum_m \vert \Psi_{0,m}\vert^2 = b/2\pi\,.
$$
Thus the electric field increases the density but independent of the distance
from the origin though the more distant Larmor circles are pulled further
apart (Fig.1).
\end{enumerate}

\medskip

\section{Second quantization}

In first quantization, the Hilbert space splits into a tensor product and
so does the observable algebra $\cO = \{a\}\otimes\{c\}$. The first factor
corresponds to the Larmor motion and the second one to the Hall
current. Upon second quantization we similarly get a tensor product of
two $(1\!+\!1)$-dimensional field theories. For small $E/B$ the first one
describes the microscopic Larmor circles while the macroscopic Hall
current is given by the second one which we shall now investigate
more closely.

To come to the many body aspects we start with a Fermi-field $\psi(x)$,
\beq
\{\psi(x), \psi^*(x')\} = \delta(x-x')\,, \qquad
\{\psi(x), \psi(x')\} = 0\,, \quad x\in{\bf R}^2
\eeq
and construct creation anf annihilation operators for the various modes by
$\int d^2 x \,\Psi^*_{n,m}(x) \psi(x)$ and the hermitian conjugate. We shall
start with $2M+1$ modes in the lowest Landau level $\psi_{0,m}$ and
consider the limit $M\ra\infty$, $E\ra 0$, such that $(2M+1)v$ stays
 less than $(b+B)/2$ in order  not to cross the next Landau level. For
our result it is essential that a finite fraction of the modes are filled.
Defining
\beq
a_{-M+m} := \int dx \,\Psi^*_{0,m}(x) \psi(x)
\eeq
we thus embed our operators in the CAR-algebra $\A$ generated by $a_m,
a_m^*$, $m\in{\bf Z}$,
\beq
\{a_m, a_n^*\} = \delta_{mn}\,, \qquad \{a_m, a_n\} = 0
\eeq
and normalize the chemical potential to zero by using the ground state
\beq
\ba{rcl}
\langle a_m^* a_n \rangle & = & \delta_{mn}\Theta(-m)\\[6pt]
\langle a_n a_m^* \rangle & = & \delta_{mn}\Theta(m)\,,
\ea
\eeq
$\Theta$ being the step function (with $\Theta(0)=1/2$). For us the
relevant observable will be the Hall current as a function of the rotation
angle $\theta$. Thus we introduce the fields
\beq
\psi(\theta) = \sum_{m} a_m e^{im\theta} \Theta(M-|m|)
\eeq
and the current
\beq
j_M(\theta) = \psi^*(\theta)\psi(\theta) -
\langle \psi^*(\theta)\psi(\theta) \rangle \,.
\eeq

So far, these operators are bounded but for $M\ra\infty$ they become
operator valued distributions and to get operators in this limit we have
to smear them with (real) test functions, however in the discrete case this
might well be the  corresponding Kronecker $\delta$,
$$
\ba{lcl}
j_{p,M} & = & \dint_{-\pi}^\pi d\theta j_M(\theta) e^{ip\theta} = \\[6pt]
& = & \dsum_{n} :a^*_{n+p} a_n: \Theta(M-|n+p|)\Theta(M-|n|)\,,
\ea
$$
$$
:a^*a: = a^*a - \langle a^* a\rangle \,.
$$

That the limit $M\ra\infty$ makes sense is shown by

\paragraph{Lemma}(1)

{\it If $f'\in L^2$ then $j_{p,M}$ converges for $M\ra\infty$ to
some $j_p$ in the strong resolvent sense.}

\paragraph{Proof}

Strong resolvent convergence means that $j_{p,M}$ converges
strongly on a dense set of essential selfadjointness, that is we
have to show that for the corresponding vectors $|d\rangle$,
$\forall \ve>0$ there exists $N\in{\bf Z}_+$ such that
$\Vert (j_{p,M}-j_{p,M'}) |d\rangle \Vert <\ve$ $\forall M,M'>N$.
Since the strong convergence on infinitely many vectors is
awkward to demonstrate we make a detour. The KMS-state
\beq
\langle a^*_m a_n\rangle = {\delta_{mn}\over 1+e^{\beta m}} =:
\delta_{mn}\Theta_\beta(-m)
\eeq
gives in the GNS-representation $\pi_T$, $T=1/\beta$, a cyclic and
separating vector $|T\rangle$ and strong convergence on $|T\rangle$
implies strong convergence on the dense set $a|T\rangle$ if
$\Vert \tau_i(a)|\Omega\rangle\Vert < \infty$. In our case,
$$
\ba{ccl}
\Vert (j_M - j_{M'})a|T\rangle\Vert^2 & = &
\langle T| a^*(j_M-j_{M'})^2 a|T\rangle\\[8pt]
& = & \langle T| \tau_i(a) a^*(j_M-j_{M'})^2|T\rangle
\ea
$$
and this goes to zero if $\vert (j_M-j_{M'})|T\rangle\vert \ra 0$. Thus
$\pi_T(j_M)$ converges to an operator $j_{\infty,T}$. Since matrix
elements in $\pi_T$, $\langle T| e^{-ij_f}e^{ij_{\infty,T}}e^{ij_g}
|T\rangle$ converge for $T\ra 0$ and the vectors $e^{ij_f}|T\rangle$
are total in the Hilbert space of $\pi_T$, this defines an operator
$j_\infty$ in $\pi_0$ \cite{epj} and this is the one we shall use
furtheron.

 Now if $M>M'$,
$$
j_{p,M}-j_{p,M'} = \sum_{n} :a_{n+p}^*a_{n}: \R^{MM'}_{n+p,n}
$$
with
$$
\R^{MM'}_{n+p,n} = \Theta(M-|n+p|)\Theta(M-|n|) -
\Theta(M'-|n+p|)\Theta(M'-|n|)\,.
$$
Furthermore $\langle :a_m^*a_{m'}::a_n^*a_{n'}:\rangle =
\delta_{mn'}\delta_{nm'}\Theta_\beta(n)\Theta_\beta(-m)$
and we get
$$
\ba{ccl}
\vert\langle \vert j_{p,M}-j_{p,M'}\vert^2\rangle\vert & = & \dsum_{n}
\left(1+e^{-\beta n}\right)^{-1}\left(1+e^{\beta(n+p)}\right)^{-1}
\R^{MM'}_{n+p,n}
 \\[10pt]
& \leq & \dsum_{M'-|p|}^{M-|p|}
%\left(1+e^{-\beta n}\right)^{-1}\left(1+e^{\beta(n+p)}\right)^{-1}\ra 0
\frac{1}{1+e^{-\beta n}}\cdot\frac{1}{1+e^{\beta(n+p)}}\ra 0\, .
\hfill \Box \ea
$$

The state (16) is nothing but the $T\ra 0$ limit of the KMS-state of
the shift, Eq.(19). Thus the above statement carries over to (16) since
everything is continuous for $T\ra 0$.

\medskip
Strong convergence asures that the limit of a product is the product of
the limits. In particular, the commutator of limit elements is the limit
of the commutator. Next we show that the latter is an element of the
center of the strong closure of $\pi_0(\A)\,$:

\paragraph{Lemma}(2)

{\it $\forall k$, %$\tilde f, \tilde g \in L^1$
the double commutator $[\,[j_{p,M}, j_{p',M}], a_k^*]$ converges to
zero in operator norm.}

\paragraph{Proof}

We use $[a_m^*a_{m'}, a_n^*a_{n'}] = a_m^*a_{n'}\delta_{nm'} -
a_n^*a_{m'}\delta_{mn'}$ to conclude
$$
\ba{ll}
[j_{p,M}, j_{p',M}] & = \dsum_{n}
a^*_{n+p+p'}a_n \Theta(M-|n|)\Theta(M-|n+p+p'|)\\[8pt]
& \times\, \left[\Theta(M-|n+p'|) - \Theta(M-|n+p|)\right]\,.
\ea
$$
Commuting with $a_k^*$ delets $a_n$ and $\sum_n$ and we remain with
$$
a^*_{k+p+p'}\Theta(M-|k|) \Theta(M-|k+p+p'|)
\left[\Theta(M-|k+p'|) -\Theta(M-|k+p|)\right]\,.
$$
Now $\Vert a^*_{k+p+p'}\Vert = 1$ and hence
$$
\Vert[\,[j_{p,M}, j_{p',M}], a_k^*]\Vert \leq
\left(\Theta(M-|k|-|p'|) - \Theta(M-|k|-|p|)\right)\,.
$$
The latter differs from zero for $|k|+|p'|<M$, $|k|+|p|>M$ or
$|k|+|p'|>M$, $|k|+|p|<M$, so for fixed $k, p, p'$ and $M\ra\infty$ it
goes to zero.
\hfill $\Box$

\medskip
Lemma (2) means that whenever $\pi_0(\A)''$, the weak closure of
$\pi_0(\A)$,  has a trivial center, $[j_p, j_{p'}]$ is a $c$-number and the
$j_p$'s generate a bosonic current algebra. We are in this situation but
since $[j_{p,M}, j_{p',M}]$ does not converge in norm this $c$-number
depends on the representation. It equals the limit of
$\langle [j_{f,M}, j_{g,M}]\rangle$ since $\langle \,\,\rangle$ is weakly
continuous and we arrive at

\paragraph{Theorem}(1)

{\it The operators $j_p$ obey}
$$
[j_p, j_{p'}] = -p\,\delta_{p,-p'}\,.
%[j_f, j_g] = -\sum_{p\in{\bf Z}} p\tilde f(p)\tilde g(-p)\,.
$$

\paragraph{Proof}

To calculate $\langle [j_{f,M}, j_{g,M}]\rangle$ we use (16) and the
expression from the previous proof.
%and $\langle a^*_{k+p+p'}a_k\rangle = \Theta(k)\delta_{p,-p'}$.
This leads to $\sum_k  \Theta(M-|k|) \left[\Theta(M-|k-p|) -
\Theta(M-|k+p|)\right]$. For $p>0$ this is $\,-\sum_{k=-M}^{-M+p}$
and for $p<0$ it is $\sum_{k=-M}^{-M-p}$, thus altogether $-p$.
\hfill $\Box$

\medskip
Also the two-point function of $j_p$'s can easily be deduced
$$
\ba{ccl}
\langle j_p\,j_{-p'}\rangle & = &
\dsum_{m,n} \langle : a^*_{n+p}a_n::a^*_m a_{m+p'}:\rangle\\[12pt]
& \times & \Theta(M-|n|)\Theta(M-|n+p|)\Theta(M-|m|)\Theta(M-|m+p'|) \\[6pt]
& = & \dsum_n \delta_{p,p'} \Theta(-n-p)\Theta(n)\Theta(M-|n|)
\Theta(M-|n+p|)\\[10pt]
& = & -p\,\delta_{pp'}\Theta(-p)\,.
\ea
$$

We have thus arrived at the current algebra and a ground state
\beq
\ba{l}
[j_p, j_{-p'}] = -p\,\delta_{pp'}\,,\quad
j^*_p = j_{-p}\,, \quad p\in{\bf Z}\\[6pt]
\langle j_p\,j_{-p'}\rangle = -p\,\delta_{pp'}\Theta(-p)\,.
\ea
\eeq
From this we define a density $\dot\rho = j$,
\beq
\rho(\theta) := i \sum_{p\not= 0} e^{-ip\theta -\ve|p|/2}\,
j_p / p  = \rho^+(\theta) + \rho^-(\theta)\,,
\eeq
$$
\rho^*(\theta)=\rho(\theta)\,.
$$
$\ve >0$ gives a cut-off and eventually, when it has made the various
manipulations legitimate, we let $\ve\ra 0$.  The ground state
$\,|0\rangle\,$ is defined as
\beq
\langle 0|\,c^*_p = c_p\,|0\rangle = 0\,,
\eeq
where
$$
j(\theta) = \sum_{p\geq 0}\left(e^{ip\theta}c_p + e^{-ip\theta}c^*_p\right)
e^{-\ve p/2}\sqrt p/2\pi\,.
$$

For the two-point function we thus get
\beq
\ba{ccl}
\langle \rho(\theta)\rho(\theta')\rangle & = &
\langle\rho^-(\theta)\rho^+(\theta')\rangle=
\dsum_{p>0} e^{i(\theta - \theta')p - p\ve}\, /p\\[14pt]
& = & -\ln(1-e^{i(\theta - \theta') - \ve}) =: \cS(\theta - \theta')\,.
\ea
\eeq
Now we define collective operators by
\beq
\Psi_\alpha(\theta) = e^{i\alpha\rho(\theta)}\,,\qquad
\Psi^*_\alpha(\theta) = e^{-i\alpha\rho(\theta)} =
\Psi_{-\alpha}(\theta)\,.
\eeq

The two-point function of the operators (24)  can be
calculated using (22) and Hausdorff's formula since the
commutator  $[\rho^+(\theta),\,\rho^-(\theta')]$ is a
$c$-number. For  coinciding arguments it equals $\,\ln \ve\,$
and thus
\beq
\ba{ccl}
\langle \Psi_\alpha(\theta)\Psi^*_\alpha(\theta')\rangle
& = & \langle e^{i\alpha(\rho^+(\theta)+\rho^-(\theta))}
e^{-i\alpha(\rho^+(\theta')+\rho^-(\theta'))}\rangle\\[6pt]
& = & e^{\alpha^2 \ln\ve}\langle e^{\alpha^2[\rho^-(\theta),
\rho^+(\theta')]}\rangle \, = \,
\ve^{\alpha^2}e^{\alpha^2\langle\rho^-(\theta)
\rho^+(\theta')\rangle} \\[6pt]
& = & e^{-\alpha^2\cS(0)}e^{\alpha^2 \cS(\theta-\theta')}\,.
\ea
\eeq
For the general  $n$-point function the same calculation
\cite{INT} amounts to
\beq
\ba{rcl}
\langle \Psi_{\alpha_1}(\theta_1)\Psi_{\alpha_2}(\theta_2) \ldots
\Psi_{\alpha_n}(\theta_n)\rangle & = & e^{-\dsum_{r=1}^n
\alpha_r^2 \cS(0)/2 - \dsum_{r<s} \alpha_r\alpha_s \cS(\theta_r - \theta_s)}
\\[6pt]
& = & \ve^{\,\,\dsum_{r=1}^n\alpha_r^2/2} \dprod_{r<s}\left(1 -
e^{i(\theta_r - \theta_s) - \ve}\right)^{\alpha_r\alpha_s}\,.
\ea
\eeq
To get in the limit $E\ra 0$ for the time evolution $\,\theta\ra\theta +
vt\,$, with $v$ as in Eq.(9), $v< E/B$, a finite velocity, we rescale
$\theta=vx$, $-\pi/v\leq x\leq \pi/v$. Then $\,(1 - e^{i(\theta_r -
\theta_s)-\ve}) \ra -iv(x-x')+\ve\,$ and rescaling $\Psi_\alpha$ to
$\Psi_\nu(x) = \ve^{-\nu/2}v^{\nu/2}e^{i\sqrt\nu \rho(x)}$ we get,
e.g., the $\nu$-anyonic $2n$-point function
\beq
\langle\Psi_\nu(x_1)\ldots\Psi_\nu(x_n)\Psi_\nu^*(y_1)\ldots
\Psi_\nu^*(y_n)\rangle =
\dfrac{\dprod_{k<l} (x_k-x_l)^\nu\,\dprod_{k<l}(y_k-y_l)^\nu}
{(-i)^{n\nu}\dprod_{k,l}[
(x_k-y_l+i\ve)]^\nu}\,.
\eeq
This shows that for $\nu$ odd (resp. even) the $\Psi$-fields at
different points anticommute (resp. commute), however they are not
necessarily canonical. In general, in this limit $\Psi_\nu$ and
$\Psi^*_\nu$ obey anyonic commutation relations \cite{INT}.

\paragraph{Remark}(1)

The commutation relation of the $\nu$-anyons with the local electron
charge becomes
$$
[\Psi_\nu(x), \rho(x')] = \sqrt\nu \delta(x-x')\,.
$$
Thus in this thermodynamic limit it happens that the charge generated
by $\Psi_\nu(x)$ is $\sqrt\nu\,\delta(x)$. This can be understood as follows:
The electron charge density is $j(x)$ and $e^{i\int f(x)j(x)dx}$ changes
its expectation value by $f'(x)$. Therefore if $f$ tends to zero at infinity, the
total change in the charge would be zero. However, for the $\nu$-anyon
$\Psi_\nu(x)$ the corresponding function is $f(x) = \sqrt\nu\,\Theta(x)$ and
formally it induces a charge $\sqrt\nu\delta(x)$, the opposite charge being
pushed to infinity. What happens more exactly is that for the (regularized)
smearing function $f_M(x) = \sqrt\nu\,[\Theta(x) - \Theta(x-M)]$ the unitaries
$e^{i\int f_M(x)j(x)dx}$ do not converge even weakly for $M\ra\infty$ but
the transformation they induce does \cite{epj}. The $\Psi_\nu(x)$ are
namely the ideal elements added, which generate this local gauge
transformation.

\medskip
Eq.(25) can be written as $\,\left(\Det{1\over x_k-y_l}\right)^\nu\,$ and
shows only for $\nu=1$ the truncation properties of a quasifree state. The
corresponding wave functions are given only in this case by a Slater
determinant and otherwise, as we shall show below, they are of Laughlin
type of order $\nu$.

\medskip

\section{Anyons and Laughlin states}

\paragraph{Definition}(1)

An $n$-particle state is given by
$$
|n\rangle = \int \Psi^*(x_1)\dots\Psi^*(x_n)|\Omega\rangle F(x_1,\dots,
x_n)dx_1\dots dx_n\,,
$$
its wave function is
$$
\phi(x_1,\dots, x_n) := \langle\Omega|\Psi(x_1)\dots\Psi(x_n)|n\rangle \,.
$$
$|n\rangle$ is a Slater state if $F(x_1,\dots, x_n) = \dprod_{i}f_i(x_i)$
and $\phi$ is of Laughlin type of order $\nu$, if  it is of the form
$\dprod_{i>k}(x_i-x_k)^\nu\dprod_m \Phi(x_m)$, for $0<|\Phi|<\infty \,\,
%\forall x_k
$ and $\nu$ odd.

\paragraph{Theorem}(2)

{\it For fermions of order $\nu$ a Slater state constructed with (22)
has Laughlin-type wave function of order $\nu$ for a total set of $f$'s.}

\paragraph{Remarks}(2)
\begin{enumerate}
\item Because of the anti-commutativity of the $\Psi$'s, the Slater determinant
$\Det f_i(x_j)$ gives the same state as $F$.
\item If $|\Omega\rangle$ is the vacuum then $|n\rangle = 0$ if for some $f_k$,
$\mbox{supp } \tilde f_k \subset (0, -\infty)$. However, Definition (1)
can also be used for KMS-states and then Theorem (2) holds with some
minor modification.
\end{enumerate}

\paragraph{Proof}

We take $f$'s with $\mbox{supp } \tilde f_k \subset (0, \infty)$ such that
$f(x)$ is analytic in the upper half-plane. For them,
$\left\{f_z(x)=(x-z)^{-1}, \mbox{Im }z < 0 \right\}$ is total. Then we get up
to a normalization factor
\beqan
\phi(x_1,\dots, x_n) & = &\prod_{i>j}(x_i-x_j)^\nu
\int\frac{dy_1}{(y_1-z_1)}\dots
\frac{dy_n}{(y_n-z_n)}\,\frac{\dprod_{k>l}(y_k-y_l)^\nu }
{\dprod_{k,l}(x_k-y_l+i\ve)^\nu}\\[6pt]
& = & \frac{\dprod_{l>j}(x_l-x_j)^\nu \dprod_{k>l}(z_k-z_l)^\nu}
{\dprod_{k,l}(x_k-z_l+i\ve)^\nu}\, .
\eeqan
Thus we have a Laughlin type wave function with
$$
\Phi(x) = \prod_l (x-z_l+i\ve)^{-\nu}
$$
which has the desired properties.
\hfill $\Box$

\paragraph{Remarks}(3)
\begin{enumerate}
\item For $\nu = 1$ $\phi$ is (up to a constant factor) the Slater determinant
$\Det\left(x_k^{\,j} \Phi(x_k)\right)$, for other $\nu$'s it is the
$\nu$-th power of such  a determinant.
\item For finite temperature $T=\beta^{-1}$, $\pi (x_l-x_k)$ is replaced by
$\beta\,\sh\frac{\pi(x_l-x_k)}{\beta}$ and $\Phi(x)$ --- by $\beta^{n}
\prod_{l=1}^n \sh^{-\nu}[\pi(x-z_l+i\ve)/\beta]$. By pulling out
$\prod_{l>k}(x_l-x_k)^\nu$ the rest gets a factor
$\prod_{l>k}\sh^\nu [\pi(x_l-x_k)/\beta]/(x_l-x_k)$ which is finite
and symmetric but no longer a pointwise product.
\end{enumerate}

\section{Conclusions}

We have studied the typical quantum Hall setting in the spirit of
canonical quantum theory.  The key point in our analysis is the
tensor-product structure of the theory that describes it. Upon second
quantization the one-dimensional field algebra related to the fractional
quantum Hall effect is then exhibited in the thermodynamic limit. For its
construction from some collective modes it is essential for the last Landau
level to be filled to a finite fraction. This is an anyonic algebra
\cite{epj,tmp,INT} which, in particular, contains (noncanonical) Fermi-fields
characterized by odd integer values of the statistics parameter $\nu$. Despite
of being locally anticommuting, these ``fermions" are unbounded, do not
satisfy CAR's and their correlators exhibit severe temperature dependence
\cite{INT,NI}. However, their $n$-particle wave functions at zero temperature
are of Laughlin type of order $\nu$, with a simple generalization for the
finite-temperature case \cite{IT}. Thus, a relation between the first and
second quantized pictures of the FQHE is achieved.

\section*{Acknowledgements}

We thank E. Langmann and H. Narnhofer for suggestive discussions and
useful remarks.

\smallskip
N.I. thanks the International Erwin Schr\"odinger Institute for
Mathematical Physics where the research has been performed, for
hospitality and financial support. This work has been supported
in part also by ``Fonds zur F\"orderung der wissenschaftlichen
Forschung in \"Osterreich" under grant P11287--PHY.

\end{document}